\documentclass[preprintnumbers, twocolumn,showkeys,superscriptaddress,showpacs,
nofootinbib,pre,notitlepage]
{revtex4-2}
\usepackage{bbm,pifont}
\usepackage{amssymb}
\usepackage{amsmath}
\usepackage{physics}
\usepackage{enumitem}
\usepackage{amsfonts}
\usepackage{epsfig}
\usepackage{graphicx}
\usepackage{tabularx}
\usepackage{color}
\usepackage[dvipsnames]{xcolor}
\usepackage{slashed}
\usepackage{multirow}
\usepackage[colorlinks=true, linkcolor=blue, citecolor=blue, urlcolor=blue]{hyperref}   
\usepackage{algorithm}
\usepackage{algpseudocode}

\begin{document}

\title{Efficient simulation of the generalized Langevin equation using a basis of dynamically relevant histories for memory and noise}

\author{Dmitriy Kondaurov}
\email[]{kd8466@mail.ru}
\affiliation{Russian Quantum Center, 30 Bolshoy Boulevard, building 1, Skolkovo Innovation Center territory, Moscow, 121205, Russia} 
\affiliation{Moscow Institute of Physics and Technology, Institutsky lane 9, Dolgoprudny, Moscow region, 141700, Russia}
\author{Evgeny Polyakov}
\email[]{evgenii.poliakoff@gmail.com}
\affiliation{Russian Quantum Center, 30 Bolshoy Boulevard, building 1, Skolkovo Innovation Center territory, Moscow, 121205, Russia}

\begin{abstract}
The Generalized Langevin Equation (GLE) is a cornerstone for modeling non-Markovian dynamics, but its numerical solution remains challenging due to the history-dependent convolution and the need to generate colored noise with a prescribed correlation function. We introduce a forward construction that addresses both challenges within a single framework. We represent an arbitrary trajectory as sum of relevant and irrelevant parts, where relevant part can be decomposed into a small number of characteristic history patterns that make the main contribution to future dynamics. It allows to reduce numerical cost to \(O(d^2)\) for $d$ chosen patterns instead of full history. The same construction applies to colored noise generation, where the causal filter plays the role of the memory kernel. We demonstrate the accuracy and efficiency of the method for several representative kernels, including long-range kernels with power-law tails.
\end{abstract}
\maketitle

\section{Introduction}
\label{sec:introduction}

The Generalized Langevin Equation (GLE) stands as a cornerstone of non-equilibrium statistical mechanics, providing a rigorous framework for describing systems with memory effects. Originally derived through projection operator techniques \cite{Langevin1908, Mori1965, Zwanzig1973, ma2016derivation, Zwanzig1961, Zhu2018, Hanggi1993, Hanggi1990}, the GLE takes the form:
\begin{equation}
m\dot{v}(t) = F(x(t)) -\int_{-\infty}^t M(t-\tau) v(\tau) d\tau + \xi(t)
\label{eq:gle_intro}
\end{equation}
where $v(t) = \dot x(t)$ is a velocity of a particle with mass $m$, $F(x)$ is external force, $M(t)$ is the memory kernel encoding temporal correlations, and $\xi(t)$ is a colored noise with the correlation function $\langle \xi(t)\xi(s) \rangle = N(t-s)$.

This formalism finds applications in a wide range of physical problems, including anomalous diffusion and aging in complex environments \cite{Lutz2001, Porra1996, Pottier2003}, subdiffusive transport in viscoelastic media and Brownian motion with hydrodynamic memory \cite{Goychuk2012, Franosch2011}, anomalous polymer dynamics and non-Markovian coarse-grained models of macromolecules \cite{Panja2010, li2017computing, chen2014computation, Klippenstein2021}, as well as colored-noise thermostats \cite{Ceriotti2009, Ceriotti2010}. Related applications also arise in quantum dissipative systems \cite{Weiss1999, kondaurov2026quantum, gardiner2004quantum, breuer, Grabert1988}.

A central challenge is the numerical solution of the GLE, which involves the history-dependent convolution in Eq.~\eqref{eq:gle_intro} and the generation of colored noise with a determined correlation function. Traditional numerical strategies navigate a fundamental trade-off: simplifying the non-Markovian dynamics to make them tractable while attempting to retain essential memory effects. 

One of the most straightforward strategies is the direct numerical integration of the GLE. In this approach, the history-dependent convolution integral is evaluated explicitly at each time step, often using a simple quadrature rule and a time cutoff for the memory kernel \cite{Ermak1980, Smith1990}. While conceptually simple and applicable to arbitrary memory kernels, the direct evaluation of the convolution scales quadratically with the number of time steps (\(O(N^2)\)) and requires storing the full history of the trajectory, which becomes prohibitively expensive for long-time simulations. Significant efforts have been made to optimize this direct approach \cite{LopezFernandez2006, LeimkuhlerSachs2022, Jung2017}. However, direct methods still face challenges in efficiently and accurately handling strongly non-Markovian systems, especially those with long-range or power-law memory kernels like anomalous
diffusion and low-temperature Brownian motion. 

A common approach is the Markovian embedding paradigm, where the dynamics are reformulated as a high-dimensional system of coupled Markovian stochastic differential equations. For general kernels, this reformulation is not exact and requires a separate approximation step: the memory kernel must first be represented in a tractable class, typically as a positive Prony series (a sum of exponentials) \cite{Baczewski2013, bockius2021model, bockius2026determining, Bockius2025}. An alternative route to a tractable kernel representation is to reconstruct it from trajectory data, i.e., to solve the inverse problem. Several data-driven approaches have been developed in this direction, ranging from rational approximations in the Laplace domain \cite{Lei2016} and iterative reconstruction schemes \cite{Jung2017, Klippenstein2024} to deep-learning-based kernel extraction \cite{KerrWinter2023, Russo2022}. Beyond homogeneous kernels, recent work also addresses state-dependent memory \cite{Ge2024} and ab initio parametrizations \cite{Xie2024}. These methods extract the memory kernel (or its parameters) from molecular dynamics simulations and then embed the resulting kernel into a Markovian system, so they still inherit the two-step structure---reconstruction followed by embedding. Recent refinements of this concept focus on deriving systematic memory corrections within this embedded framework. For instance, the effective mass approach yields a Markovian Langevin equation with a renormalized mass, capturing leading-order memory effects for short correlation times and systematically improvable via higher-order corrections \cite{WisniewskiLuczkaSpiechowicz2024_PRE, WisniewskiLuczkaSpiechowicz2024_Entropy}. While valuable, this methodology exemplifies the prevalent two-step, prescriptive paradigm: first, the memory kernel is approximated, fitted, or reconstructed to a specific class; second, high-performance integrators are applied to the resulting approximate, quasi-Markovian system \cite{LeimkuhlerSachs2022}. Such a separation requires a high-quality initial fit of the kernel—or a high-quality reconstruction stage—which is particularly problematic for kernels of a power-law form.

Furthermore, the need to generate the correlated noise \(\xi(t)\) presents a parallel challenge, often addressed by algorithms designed for specific correlation forms or assuming underlying auxiliary noise dynamics \cite{Mannella1989, Garcia-Ojalvo1999, Hanke2024}. In cases where the classical FDT is not satisfied, the application of standard approaches becomes challenging or even impossible \cite{Netz2023, Stella2014}.

In this paper, we propose a different approach to Markovian embedding. Instead of approximating the kernel or introducing additional variables, we analyze which patterns of past history contribute most significantly to the memory-dependent convolution. Based on this, we construct an efficient, low-dimensional basis of relevant historical patterns. Consequently, rather than needing to store the entire history, we only need to retain a few decomposition components, which are sufficient to compute the convolution to a specified level of accuracy. Furthermore, we require these decomposition coefficients to evolve in a self-consistent manner, rendering them natural auxiliary Markovian variables.

Moreover, the same construction applies to the generation of colored noise. For a stationary correlation function whose covariance matrix is positive definite, a realization of the colored noise is obtained by convolving white noise with a causal filter whose autocorrelation reproduces the target correlation function. In this representation, the sequence of independent, normally distributed random variables plays exactly the same role as the velocity trajectory, while the causal filter acts as the memory kernel.

We demonstrate that this formalism accurately reproduces results from direct integration benchmarks for several representative kernels, including long-range kernels with power-law tails relevant to anomalous diffusion, where traditional Prony series approximations converge slowly. 

The paper is organized as follows. Section~\ref{sec:model} introduces the model and formulates the problem for both the memory force and the colored noise. Section~\ref{sec:construction} presents the recursive basis method, covering the construction and convergence of the basis, its application to memory and noise, and the resulting integration scheme for the GLE. Section~\ref{sec:results} reports numerical validation for representative kernels. We conclude in Section~\ref{sec:conclusion}.

\section{Model and problem formulation}
\label{sec:model}

We consider a particle described by the GLE, Eq.~\eqref{eq:gle_intro}, where \(M(t)\) is the memory kernel characterizing the time-nonlocal friction and \(\xi(t)\) is a stationary Gaussian noise with zero mean, \(\langle\xi(t)\rangle=0\), and correlation function \(\langle\xi(t)\xi(s)\rangle = N(t-s)\); in general, \(N(t)\) is not related to \(M(t)\). 

For example, for quantum Brownian motion at zero temperature \cite{kondaurov2026quantum} these functions are $M(t) \propto \varepsilon / (\varepsilon^2+t^2)$ and $N(t) \propto (\varepsilon^2 - t^2) / (\varepsilon^2+t^2)^2$. Noise correlation function has slow negative algebraic tails and does not satisfy the classical FDT.  

The numerical challenge is evaluating the memory force and generating colored noise. We begin with the memory force which can be rewritten as

\begin{equation}
    F_{mem}(t) = \int_0^{\infty} M(s) v(t-s) ds.
\end{equation}

For its direct computation we must use whole past history, which is numerically expensive. We simplify this problem by the next way. Represent history as

\begin{equation}
    v(t-s) = v_{rel}(t-s) + \delta v (t-s).
\end{equation}

Let us assume that there exist characteristic patterns $v_i(s)$ such that $v_{rel}(t-s) = \sum_{i=1}^{r} a_i(t) u_i(s)$, where $r$ is a sufficiently small number ($\sim 5–10$) and $a_i(t)$ are the expansion coefficients. It is assumed that the functions $u_i(s)$ are known; thus, the memory force is uniquely determined solely by the coefficients $a_i(t)$. We are looking for a representation in which the full memory dynamics are governed exclusively by the self-consistent dynamics of these coefficients. 

Naturally, it is necessary to require that the irrelevant part $\delta v(t-s)$ makes a controllably negligible contribution to the memory force, and that it can be disregarded without introducing errors at future times.

\subsection{Problem statement}

We discretize the dynamics with a uniform time step \(\Delta t\), using a past window \(t_{\text{past}} = p\,\Delta t\) and a future window \(t_{\text{future}} = f\,\Delta t\), with positive integers \(p\) and \(f\). Both windows must be much larger than the memory time \(\tau_M\), so that the contribution to the integral beyond them is negligible.

At time step \(n\), let \(\mathbf{x}_n \in \mathbb{R}^{p}\) be the velocity history, \((\mathbf{x}_n)_j = v(t_n - j\Delta t)\) for \(j = 0, \ldots, p-1\). The memory kernel is discretized as \(m_i = \Delta t\, M(i\Delta t)\) for \(i = 0, \ldots, p-1\), and \(m_i = 0\) for \(i \geq p\), reflecting the truncation of the kernel beyond the past window. The memory force at time \(t_n\) is then
\begin{equation}
F_n^{\text{mem}} = -\mathbf{m}^\top \mathbf{x}_n = -\sum_{j=0}^{p-1} m_j\,(\mathbf{x}_n)_j.
\label{eq:memory_force_discrete}
\end{equation}

At each time step, the memory force requires evaluating the dot product $\mathbf{m}^\top\mathbf{x}_n$, which costs $p$ operations. We a looking for a low-dimensional subspace $\mathcal{V}\subset\mathbb{R}^p$ of dimension $r\ll p$ that contains nearly all the information about any possible history in its expansion coefficients.

Formally, let $V\in\mathbb{R}^{p\times r}$ be an orthonormal matrix, $V^\top V=I_r$, and let $\mathcal{V}=\operatorname{range}(V)$. The coordinate map is
\begin{equation}
    \mathbf{a}_n = V^\top\mathbf{x}_n \in \mathbb{R}^r.
\end{equation}

We want to find $V$ and the dynamics operator $D\in\mathbb{R}^{r\times(r+1)}$ such that
\begin{align}
    F_n = -\mathbf{m}^\top\mathbf{x}_n &\approx -\mathbf{m}_{rel}^\top\mathbf{a}_n,
    \label{eq:req-force}\\
    \mathbf{a}_{n+1} &= D\begin{pmatrix} v_{n+1}\\ \mathbf{a}_n\end{pmatrix},
    \label{eq:req-dynamics}
\end{align}
where $\mathbf{m}_{rel}:=V^\top\mathbf{m}$. The first requirement ensures a correct calculation of the memory force: $\mathbf{a}_n$ contains all the necessary information about the full history. This is a static condition for the relevance of the selected subspace. The second ensures dynamical consistency: the evolution of $\mathbf{a}_n$ is determined only by $\mathbf{a}_n$ and the most recent velocity $v_{n+1}$.

To formalize these requirements, we split the history as
\begin{equation}
    \mathbf{x}_n = V\mathbf{a}_n + \delta \mathbf{x}_n,
    \qquad
    \delta \mathbf{x}_n := (I - VV^\top)\mathbf{x}_n,
    \label{eq:history_split}
\end{equation}
so that the memory force becomes
\begin{equation}
    F_n = -\mathbf{m}^\top\mathbf{x}_n
    = -\underbrace{\mathbf{m}^\top V}_{=\mathbf{m}_{rel}^\top}\mathbf{a}_n
    - \mathbf{m}^\top\delta \mathbf{x}_n.
    \label{eq:exact-force}
\end{equation}

Let’s examine both requirements in turn.

\subsection{Relevant subspace and memory intensity operator}

The requirement \eqref{eq:req-force} is therefore equivalent to $\mathbf{m}^\top\delta \mathbf{x}_n\approx 0$: discarded, irrelevant part does not contribute into the memory force. However, we must also ensure that it does not contribute in the future.

The full history evolves as
\begin{equation}
    \mathbf{x}_{n+1} = S\mathbf{x}_n + v_{n+1}\mathbf{e}_{\text{in}},
    \label{eq:original_history_evolution}
\end{equation}
where $S$ is the shift operator and $\mathbf{e}_{\text{in}} = [1,0,\ldots,0]^\top$ is a unit vector representing an injection concentrated at the most recent past. Substituting the split \eqref{eq:history_split} into \eqref{eq:original_history_evolution} and iterating $s$ times, we obtain
\begin{equation}
    \mathbf{x}_{n+s} = S^s\mathbf{x}_n + \sum_{k=1}^{s}v_{n+k}S^{s-k}\mathbf{e}_{\text{in}},
\end{equation}
and hence
\begin{equation}
    F_{n+s} = -\bigl(\mathbf{m}^{(s)}\bigr)^\top\mathbf{x}_n - \sum_{k=1}^{s} m_{s-k}\,v_{n+k},
    \label{eq:F-ns}
\end{equation}
where $\mathbf{m}^{(s)}_j = m_{s+j}$, with the convention $m_i=0$ for $i\ge p$. The residual contribution to the future force is $(\mathbf{m}^{(s)})^\top\mathbf{r}_n$, so the requirement $\mathbf{m}^\top\mathbf{r}_n\approx 0$ extends to
\begin{equation}
    \bigl|(\mathbf{m}^{(s)})^\top\mathbf{r}_n\bigr|\approx 0
    \quad \text{for all } s=0,\ldots,f-1.
\end{equation}
Combining these requirements, we obtain final static condition for relevant basis
\begin{equation}
    E(V) = \sum_{s=0}^{f-1}\bigl\|\mathbf{m}^{(s)} - VV^\top\mathbf{m}^{(s)}\bigr\|^2 \xrightarrow[]{}\min.
    \label{eq:E-functional}
\end{equation}

We introduce the memory intensity operator $R\in\mathbb{R}^{p\times p}$, which aggregates the influence of the history over all future shifts within the considered window:
\begin{equation}
    R := \sum_{s=0}^{f-1} \mathbf{m}^{(s)}\mathbf{m}^{(s)\top}.
    \label{eq:R-def}
\end{equation}
In component form, $R_{ij} = \sum_{s=0}^{f-1} m_{s+i} m_{s+j}$, with the convention $m_i=0$ for $i\ge p$. 

For a unit-norm history pattern $\mathbf{x}$, the quadratic form $\langle\mathbf{x},R\mathbf{x}\rangle$ measures the total squared force exerted on the future by this pattern over the future window. Thus, \eqref{eq:E-functional} is equivalent to 
\begin{equation}
    E(V) = \sum_{s=0}^{f-1}\bigl\|\mathbf{m}^{(s)}\bigr\|^2 - \operatorname{tr}\bigl(V^\top R\,V\bigr).
    \label{eq:E-expansion}
\end{equation}

\subsection{Dynamical consistency}

Except for a basis \(V\in\mathbb{R}^{p\times r}\) that approximately satisfies the static requirement \eqref{eq:E-expansion} we also need the dynamics operator with the dynamical consistency condition introduced below. 

The simulating dynamics requires us to track the system's position within a subspace as new information arrives at each time step. Consider the system at a discrete time step \(n\). We maintain a set of \(r\) orthonormal basis vectors, stored as columns in a matrix \(V\in\mathbb{R}^{p\times r}\). Each column represents a memory pattern that is currently relevant.

As time advances by one step \(\Delta t\), two operations occur. Firstly, aging --- all current memory patterns shift one step into the past. Mathematically, this is equivalent to shifting the basis, \(S(V)\). Second,  injection --- a new piece of information enters the system: the velocity \(v_{n+1}\) at the next time step. This corresponds to the unit vector \(\mathbf{e}_{\text{in}}\) in the history space.

Dynamical consistency requires that if \(\mathbf{x}_n\in\mathcal{V}\) (so \(\mathbf{x}_n\approx V\mathbf{a}_n\)), then the updated history \(\mathbf{x}_{n+1}=SV\mathbf{a}_n+v_{n+1}\mathbf{e}_{\text{in}}\) also remains in \(\mathcal{V}\). In matrix form, the ideal target is
\begin{equation}
U \begin{pmatrix}v_{n+1}\\ \mathbf{a}_n\end{pmatrix}
\approx
VD
\begin{pmatrix}v_{n+1}\\ \mathbf{a}_n\end{pmatrix},
\end{equation}
where $U = \begin{pmatrix}\mathbf{e}_{\text{in}} & SV\end{pmatrix}$. Since the pair \((v_{n+1},\mathbf{a}_n)\) is arbitrary, this yields the dynamical condition
\begin{equation}
U \approx VD.
\label{eq:dynamic1}
\end{equation}
Equation \eqref{eq:dynamic1} for \(r<p\) cannot hold exactly. However, we do not require this equality to hold strictly. It suffices for the difference between the left- and right-hand sides of the equation to lie within an irrelevant space. Mathematically, this is equivalent to the projection of the residual onto the memory intensity matrix $R$ is minimal. Denote $\delta U = U - VD \in \mathbb{R}^{p \times (r+1)}$, so, by analogy with \eqref{eq:E-expansion}, we obtain the second functional

\begin{equation}
    \operatorname{tr}\bigl(\delta U^\top R\, \delta U\bigr)\ \text{is minimized}.
    \label{eq:dynamic}
\end{equation}

Simultaneously, the static requirement must be satisfied:
\begin{equation}
    \operatorname{tr}\bigl(V^\top R\,V\bigr)\ \text{is maximized}.
    \label{eq:static}
\end{equation}

In summary, we are looking for a subspace that is memory-intensity-maximizing and the evolution rule which maintain this subspace approximately shift-invariant in the sense of memory intensity. We choose a basis in the history space such that each history splits into relevant and irrelevant parts, where the relevant part contains all the information necessary to calculate the memory force and its own evolution \eqref{eq:dynamic}, while the irrelevant part makes no contribution to future \eqref{eq:static}.

\subsection{Colored noise: from correlation function to causal filter}
\label{sec:causal_filter}

To generate noise with a specified correlation \(N(t)\), we are looking for a causal operation that maps an uncorrelated Gaussian white noise sequence \(\{\zeta_n\}\) into the desired colored noise \(\{\xi_n\}\). Specifically, we want to find a causal filter \(\mathbf{c} = [c_0, c_1, \ldots, c_{p-1}]^\top\) such that the convolution
\begin{equation}
\xi_n = \sum_{j=0}^{p-1} c_j \, \zeta_{n-j}
\label{eq:noise_convolution}
\end{equation}
reproduces the target correlations:
\begin{equation}
\langle \xi_n \xi_{n+k} \rangle = N(k\Delta t) \quad \text{for } k = 0,1,\ldots,p-1.
\label{eq:filter_reconstructs_kernel}
\end{equation}
The filter coefficients must satisfy
\begin{equation}
\sum_{j=0}^{p-k-1} c_j c_{j+k} = N(k\Delta t) \quad \text{for } k = 0,1,\ldots,p-1.
\label{eq:filter_coefficients}
\end{equation}
Thus the filter encodes the same correlation structure, represented in a causal form suitable for sequential noise generation.

Finding this filter is equivalent to computing the square root of the noise covariance matrix \(T\), where \(T_{ij} = N(|i-j|\Delta t)\); we assume $T$ is positive definite. The Cholesky decomposition \(T = L L^\top\) provides such a factorization, with \(L\) lower-triangular. For a stationary correlation \(N(t)\), the columns of \(\L\) become approximately translationally invariant far from the matrix boundaries: \(L_{i+k,j+k}\approx L_{i,j}\) for sufficiently large \(k\).

The causal filter \(\mathbf{c}\) can be treated as a memory kernel acting on past white noise, so the problem of generating colored noise reduces entirely to the problem of calculating the memory force.

\section{Basis construction and GLE integration}
\label{sec:construction}

In this section we discuss explicit and recursive approaches to construction of the basis that satisfies the given requirements \eqref{eq:dynamic} and \eqref{eq:static}.

\subsection{Explicit construction of relevant subspace}

Let's consider the spectral decomposition of $R$:
\begin{equation}
R \mathbf{u}_i = \lambda_i \mathbf{u}_i, \quad \mathbf{u}_i \in \mathbb{R}^{p}, \quad \lambda_1 \geq \lambda_2 \geq \cdots \geq \lambda_{p} \geq 0.
\label{eq:eigenproblem}
\end{equation}
The eigenvectors $\mathbf{u}_i$ form an orthonormal basis for the space of past histories. Their corresponding eigenvalues $\lambda_i$ quantify the importance of each eigenvector: a history aligned with $\mathbf{u}_i$ couples to the future with an intensity proportional to $\lambda_i$.

By Ky Fan's maximum principle \cite{Fan1949}, $\operatorname{tr}(V^\top R V)$ is maximized when the columns of $V$ are the eigenvectors of $R$ corresponding to the $r$ largest eigenvalues: $V = (\mathbf{u}_1 \ldots \mathbf{u}_r)\in\mathbb{R}^{p\times r}$. Thus, the maximally-static-optimal basis for a fixed dimension r is 
\begin{equation}
\mathcal{S}_{\text{rel}} = \{ \mathbf{u}_1 \ldots \mathbf{u}_r  \},
\label{eq:relevant_states_def}
\end{equation}

The span of $\mathcal{S}_{\text{rel}}$ constitutes the relevant subspace $\mathcal{V} \subset \mathbb{R}^{p}$, a low-dimensional subspace that captures the dominant pathways through which the past influences the future for the given kernel $M(t)$.

\begin{figure}[t]
    \centering
    \includegraphics[width=\columnwidth]{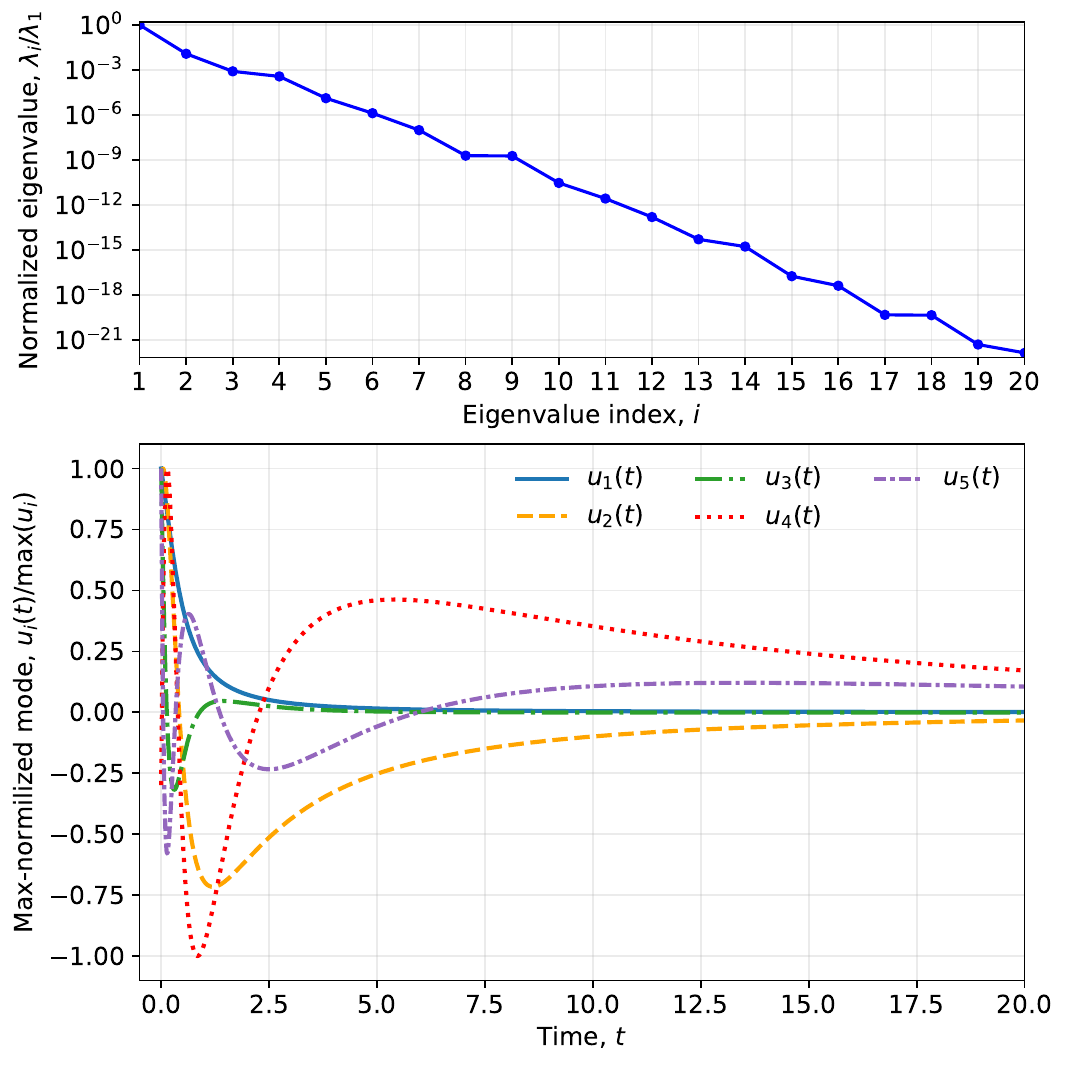}
    \caption{Spectrum and first eigenfunctions of the memory intensity operator $R$ for the memory function $M(t) = \tau / (t^2 + \tau^2)$. The eigenvalues $\lambda_i$ decay rapidly with the mode index $i$, indicating that only the first few modes carry significant weight in the action of the kernel.}
    \label{fig:spectrum_R}
\end{figure}

The spectrum of $R$ for a representative kernel is shown in Fig.~\ref{fig:spectrum_R}: the eigenvalues decay rapidly with the mode index, indicating that only the first few modes carry significant weight in the action of the kernel. These few modes are also shown in the bottom plot. 

If we choose $V$ by this way, we can also explicitly obtain dynamics operator. For fixed $V$, the condition \eqref{eq:dynamic} is least squares problem and it has solution \cite{Wei2000}

\begin{equation}
    D = (V^\top R V)^{-1} V^\top R U 
\end{equation}

For $V$ which composed from eigenvectors of $R$ this formula is simplified to

\begin{equation}
    D = V^\top U 
\end{equation}

\subsection{Recursive construction}
\label{sec:recursive_algorithm_A}

Other approach is recursive construction. We start from static-optimal $V_0 = (\mathbf{u}_1 \ldots \mathbf{u}_r)$. Suppose, that on the step $n$ we have any $V_n$. So on the step $n+1$ we construct an augmented basis $U_{n+1}$ by concatenating the new information with all the aged patterns from the previous step:
\begin{equation}
U_{n+1} = (\mathbf{e}_{\text{in}}\; S V_n ) \in \mathbb{R}^{p \times ({r+1})}.
\label{eq:augmented_basis}
\end{equation}
This collection of $(r+1)$ vectors spans a space that contains all possible linear combinations of the fresh data and aged memories. The dimensionality has temporarily increased by one.

The augmented basis $U_{n+1}$ is overcomplete. To maintain a fixed, minimal dimension $r$, we must compress this information back into the $r$-dimensional relevant subspace $\mathcal{V}$ defined by the memory kernel. This is achieved by projecting the memory intensity operator $R$ onto the augmented basis:
\begin{equation}
\rho_{n+1} = U_{n+1}^\top R \, U_{n+1} \in \mathbb{R}^{(r+1) \times (r+1)}.
\label{eq:rho_matrix}
\end{equation}
The matrix $\rho_{n+1}$ encodes how strongly each vector in $U_{n+1}$ (and their combinations) couples to future dynamics through the memory kernel.

We then perform an eigendecomposition of $\rho_{n+1}$:
\begin{equation}
\rho_{n+1} \widetilde{U}_{n+1} = \widetilde{U}_{n+1} \Pi_{n+1},
\end{equation}
where $\Pi_{n+1}$ is a diagonal matrix of eigenvalues $\pi_1 \ge \pi_2 \ge \dots \ge \pi_{r
+1} \ge 0$, and $\widetilde{U}_{n+1}$ contains the corresponding orthonormal eigenvectors. These eigenvectors represent the optimal linear combinations of the fresh and aged data, ordered from most to least important in terms of their future influence.

The optimal combinations are mapped back to the original history space to form a new, full set of candidate patterns:
\begin{equation}
V_{n+1}^{\text{(full)}} = U_{n+1} \, \widetilde{U}_{n+1} \in \mathbb{R}^{p \times (r+1)}.
\label{eq:full_new_basis}
\end{equation}
The columns of $V_{n+1}^{\text{(full)}}$ are orthonormal and ranked by importance. To complete the recursion and maintain a fixed state dimension, we discard the least important pattern—the one corresponding to the smallest eigenvalue $\pi_{r+1}$:
\begin{equation}
V_{n+1} =  V_{n+1}^{\text{(full)}} (I_r \, \, \, \mathbf{0})^\top \in \mathbb{R}^{p \times r}.
\label{eq:truncated_basis}
\end{equation}
The matrix $V_{n+1}$ now contains our updated orthonormal basis for time step $n+1$. 
Iterating this procedure from a suitable initial basis $V_0$ (e.g., the dominant eigenvectors of $R$) leads to convergence. The sequence $\{V_n\}$ typically reaches a steady state $V_{\infty}$, where the update becomes a pure rotation within the relevant subspace $\mathcal{V}$.
The associated transformation matrix $\widetilde U_{\infty}$ is used in the dynamical integration:

\begin{equation}
    D := (I_r \, \, \, \mathbf{0}) \widetilde U_\infty^\top. 
\end{equation}

\subsection{Colored Noise Generation}
\label{sec:colored_noise}

By Sec.~\ref{sec:causal_filter}, we have reduced the problem of generating colored noise to the problem of finding a retarded force. We proceed exactly as in Sec.~\ref{sec:model} and \ref{sec:construction}, but now with $\mathbf{c}$ playing the role of the discretized kernel. We construct the memory intensity operator $R_C$ for the filter:
\begin{equation}
R_C = \frac{1}{f}\sum_{s=0}^{f-1} \mathbf{c}^{(s)} (\mathbf{c}^{(s)})^\top,
\end{equation}
where $\mathbf{c}^{(s)} = [c_s, \ldots, c_{s+p-1}]^\top$ is a segment of the filter. 
Repeating the procedure in its entirety for the memory we obtain the relevant basis and dynamics operator for the noise process.

\subsection{Solving the GLE with the Recursive Basis}
\label{sec:solving_gle}

The final algorithm combines both tasks. We maintain two state vectors in the relevant subspaces $\mathcal{V}_M$ and $\mathcal{V}_N$: $\mathbf{a}_n$ -- the deterministic memory state, compressing past velocities and $\mathbf{b}_n$ -- the stochastic noise state, compressing past white noise inputs.

\begin{figure}[h]
    \centering
    \includegraphics[width=0.5\textwidth]{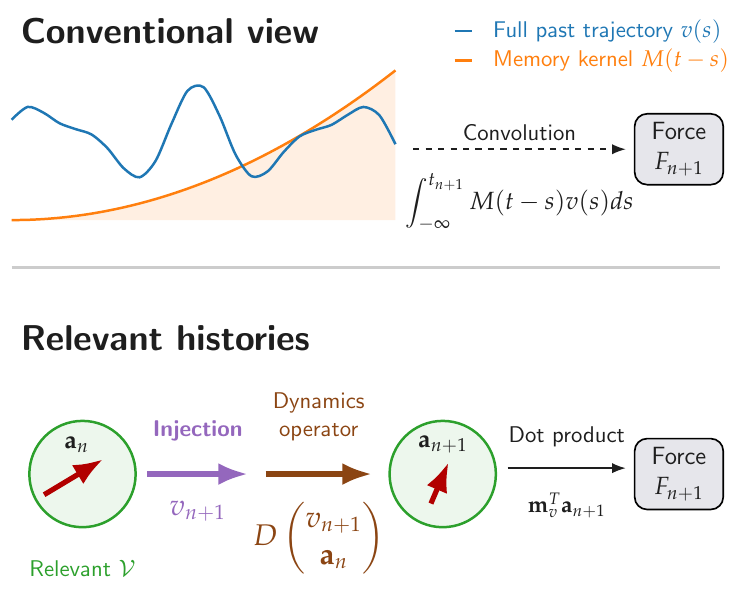}
    \caption{\textbf{Schematic of the relevant histories approach.} }
    \label{fig:scheme}
\end{figure}

Recall the discretized GLE for a particle of unit mass, driven by a colored noise $\xi_n$:
\begin{equation}
v_{n+1} - v_n = \Delta t (F(x_n) + F_{mem, n} + \xi_n).
\label{eq:gle_direct_discrete}
\end{equation}

Let us describe the complete algorithm. During the preliminary computation stage, we calculate a causal filter for the noise and, using either an explicit or a recursive construction, form the corresponding bases $V$ and $V^{(N)}$, as well as the dynamics operators $D$ and $D^{(N)}$ — for both memory and noise.

At each time step we calculate the memory force as a simple low-dimensional dot product

\begin{equation}
F_{mem, n}  \approx \mathbf{m}_{rel}^\top \mathbf{a}_n, \quad \text{where} \quad \mathbf{m}_{rel} = V_{}^\top \mathbf{m}.
\label{eq:convolution_approximation}
\end{equation}

The noise $\xi_n$ defined by correlation function $C(t)$ is similarly constructed as

\begin{equation}
\xi_n  \approx \mathbf{c}_{rel}^\top \mathbf{b}_n, \,\,\, \text{where} \,\,\,\, \mathbf{c}_{rel} = V_{}^{(N)\top} \mathbf{c}
\end{equation}

For the next time step, we update the relevant vectors

\begin{equation}
     \mathbf{a}_{n+1} = D^{}\begin{pmatrix} v_{n+1}\\ \mathbf{a}_n\end{pmatrix}
\end{equation}
\begin{equation}
     \mathbf{b}_{n+1} = D^{(N)}\begin{pmatrix} \zeta _{n+1}\\ \mathbf{b}_n\end{pmatrix}
\end{equation}

This scheme is visualized in Fig.~\ref{fig:scheme}.

\section{Results}

\label{sec:results}

In this section we provide the applicability of our method for some interest cases which especially can be difficult for other established approaches. We compare the recursive method with these approaches.

\subsection{Logarithmic diffusion of quantum Brownian particle at zero temperature}

\label{logdiffusion}

We compare the numerical simulation by our method with analytical result for quantum Brownian particle in the Ohmic bath at $T=0$ \cite{Grabert1988, kondaurov2026quantum}. 
\begin{equation}
    M(t) = \frac{2 }{\pi} \frac{\varepsilon}{\varepsilon^2 + t^2}
\end{equation}
\begin{equation}
    N(t) = \frac{1}{\pi} \frac{\varepsilon^2 - t^2}{(\varepsilon^2 + t^2)^2}
\end{equation}

\begin{figure}[h]
    \centering
    \includegraphics[width=\columnwidth]{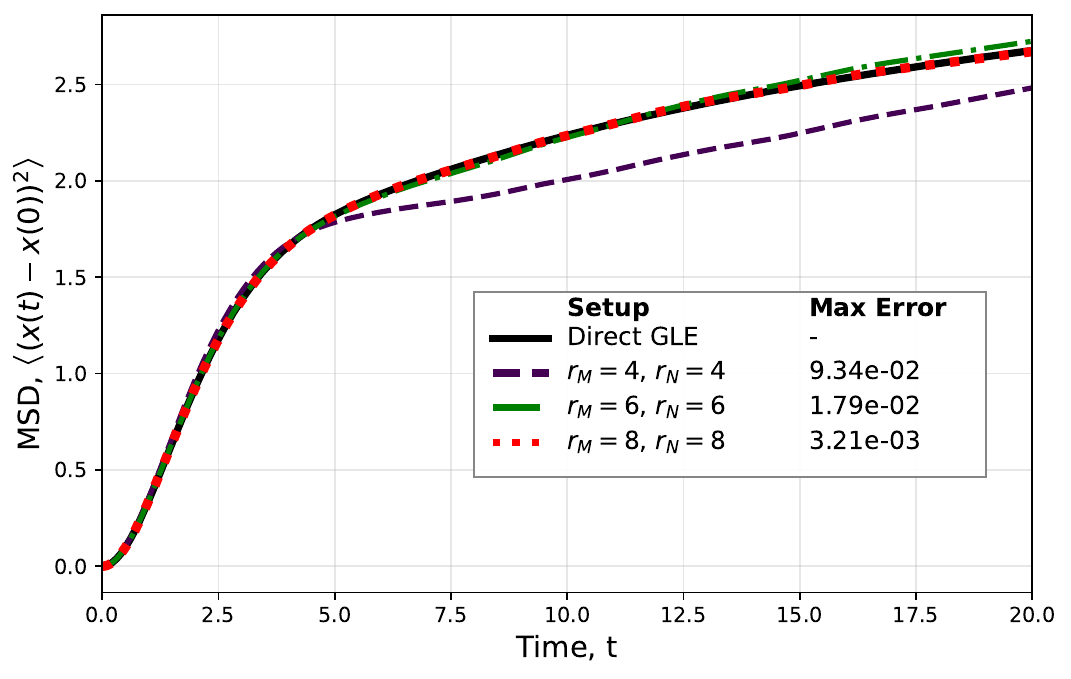}
    \caption{Comparison of the simulated equilibrium mean squared displacement (MSD) $\langle (x(t) - x(0))^2 \rangle$ with the direct GLE integration for different $r$ for quantum Brownian motion at $T=0$. Parameters: $\varepsilon = 0.5$, $m=1$, $dt=0.001$, $N_{traj} = 2 \cdot 10^5$.}
    \label{fig:msd_comparison}
\end{figure}

In Fig. \ref{fig:msd_comparison} the equilibrium mean square displacement is shown for $\varepsilon = 0.5$ for different $r$. It is essentially non-Markovian dynamics with long algebraic tails, but already 8 numbers for both of memory and noise instead of whole history provide almost true result with error $\sim 0.3 \%$.  

\subsection{Subdiffusion}

\begin{figure}[h]
    \centering
    \includegraphics[width=\columnwidth]{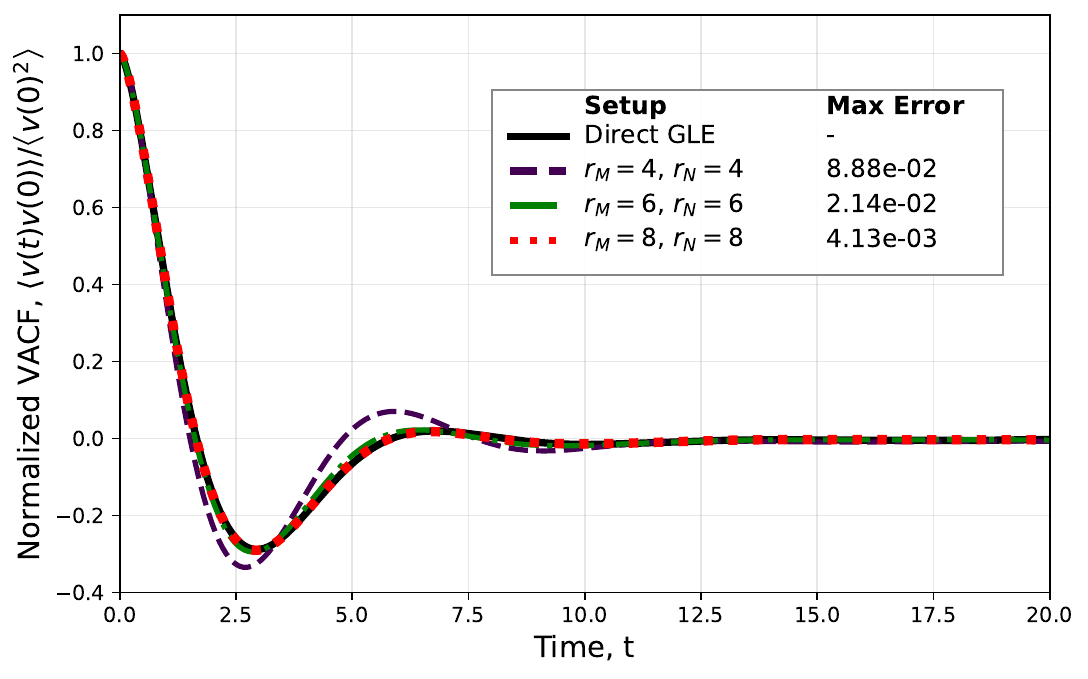}
    \caption{Comparison of the simulated velocity autocorrelation function (VACF) $\langle v(t)v(0) \rangle$ with the direct GLE integration for different $r$ for subdiffusion. Parameters:  $m=1$, $dt=0.01$, $N_{traj} = 2 \cdot 10^5$.}
    \label{fig:subdiffusion}
\end{figure}

Another interesting problem is classical subdiffusion ($N(t) = M(t) = 1 / \sqrt{\pi t}$) \cite{bockius2021model, vinales2006anomalous}. This case is quite challenging to investigate using standard methods. Fig.~\ref{fig:subdiffusion} shows the velocity autocorrelation function (VACF) for various values of $r$. When using $8+8$ relevant modes, the maximum relative error is approximately $0.4\%$.

\subsection{Comparison with other Markovian embedding approaches}

A comparison was made with the Markovian embedding method \cite{Baczewski2013}, which utilizes a positive Prony representation. For a subdiffusive kernel ($M(t) = 1/\sqrt{\pi t}$), the computational efficiency is similar to that of the relevant histories method: given a comparable number of auxiliary variables, both approaches yield a similar level of accuracy for the VACF. However, the accuracy—and, consequently, the exact number of auxiliary modes required—depends on the specific method used to approximate the memory function, a matter that warrants further analysis. The relevant histories method does not require such an analysis.

In the case of quantum Brownian motion at $T=0$, the original algorithm cannot be applied directly, as it relies on the FDT and a positive Prony representation—conditions that are not met here, given the specified memory and noise correlation. 

\section{Conclusion}

\label{sec:conclusion}

Numerical simulation based on the generalized Langevin equation (GLE) requires the computation of a history-dependent memory term, as well as the generation of colored noise with a specified correlation function—both of which are highly computationally intensive tasks. While existing approaches have significantly alleviated these challenges, they still have inherent limitations. Advanced direct integration methods accelerate the calculation of the convolution; however, because they remain direct integration techniques, the computational cost increases substantially as the time step is reduced or when dealing with long-memory kernels. Markovian embedding methods, such as those based on Prony series, allow for a representation of the system in a lower-dimensional space; yet, they exhibit slow convergence for kernels with algebraic tails and cannot account for kernels that are not positive-definite. Furthermore, such methods are often strictly tied to the classical FDT, which limits their applicability to scenarios where the noise does not satisfy this theorem. Consequently, there is a need for a method capable of handling kernels with algebraic tails (without relying on exponential approximation), accommodating kernels that are not positive-definite, generating noise independently of the FDT, and maintaining linear computational complexity with respect to the number of simulation steps.

We propose a method that addresses these limitations by analyzing which patterns of the past history contribute most significantly to the memory convolution. Instead of approximating the memory kernel, we represent the history in a low-dimensional basis of relevant patterns, selected so as to capture the dominant contribution to the convolution. By additionally requiring dynamical consistency—that the expansion coefficients evolve in a closed, self-contained manner—we obtain a natural Markovian embedding: the coefficients become auxiliary variables whose evolution is determined solely by themselves and the current velocity, and whose cost per step is far below that of direct convolution. 

Furthermore, we represent noise using a causal filter, which allows the problem of noise generation to be reduced entirely to the calculation of the memory strength. This approach does not require the assumption of positivity or the classical FDT, enabling the modeling of a much broader class of noise—a capability that is particularly essential for non-equilibrium molecular dynamics or low-temperature quantum Brownian motion.

This approach offers several key advantages:
\begin{itemize}
    \item It provides a direct construction that avoids kernel fitting and the introduction of auxiliary variables, while keeping the full, non-Markovian GLE as the starting point.
    
    \item It delivers a controlled compression of the memory: for a given tolerance, the relevant basis retains the dominant spectral components of the memory operator and provides a direct measure of the effective dimensionality of the memory process.
    
    \item It reduces the cost of history integration from \(O(N^2)\) to \(O(d^2 N)\) for a \(d\)-dimensional basis, so that the cost scales linearly with the number of simulation steps for a fixed kernel.

    \item It accommodates kernels that are not positive definite---including alternating-sign kernels, for which a positive Prony representation is not available---without requiring a different algorithmic structure, and it applies to noise correlation functions that are not tied to the kernel by the classical fluctuation-dissipation theorem.
    
\end{itemize}

We test our method on challenging benchmarks—subdiffusion and quantum Brownian motion at $T=0$—where the Prony representation is inapplicable and the classical FDT does not hold. Nevertheless, in these cases, a small number of relevant modes provide good agreement with direct integration.

In summary, we propose a natural Markovian embedding method, in which the auxiliary variables are defined directly from the structure of the kernel. This approach allows for the efficient calculation of memory strength and the generation of colored noise with a specified correlation, which significantly facilitates the simulation of GLE with challenging kernels.

\bibliography{main}

\end{document}